\documentclass[a4paper,12pt]{article}
\pdfoutput=1
\usepackage{feynmp-auto,expdlist}
\usepackage{amsmath,amsfonts,amssymb}
\usepackage{graphicx}
\usepackage{enumerate}
\usepackage{hyperref}
\usepackage{latexsym}
\usepackage{dsfont}
\usepackage{hepnicenames}
\usepackage{enumerate}
\usepackage{soul}
\usepackage[normalem]{ulem}
\usepackage{amsmath}
\usepackage{amssymb}
\usepackage{makecell}
\usepackage{bbm}

\font\tenrsfs=rsfs10 at 12pt
\font\sevenrsfs=rsfs7
\font\fiversfs=rsfs5
\newfam\rsfsfam
\textfont\rsfsfam=\tenrsfs 
\scriptfont\rsfsfam=\sevenrsfs
\scriptscriptfont\rsfsfam=\fiversfs

\numberwithin{equation}{section}

\newcommand\numberthis{\addtocounter{equation}{1}\tag{\theequation}}

\usepackage{mathrsfs}
\usepackage{braket}
\usepackage{titling}
\usepackage{amsmath}
\usepackage{slashed}
\usepackage{amssymb}
\usepackage{epsfig}
\usepackage{graphicx}
\usepackage{color}
\usepackage{rotating}
\usepackage{hyperref}
\usepackage[table,xcdraw,dvipsnames]{xcolor}
\usepackage[compress,numbers,sort]{natbib}
\usepackage{colortbl}
\usepackage{pdflscape}
\usepackage{color}
\usepackage{mathtools}
\usepackage{colortbl}
\definecolor{Gray}{gray}{0.95}
\definecolor{RGray}{gray}{0.85}
\definecolor{CGray}{gray}{0.93}

\newcommand{\G}{{\cal G}}
\renewcommand{\P}{{\cal P}}

\newcommand{\M}{{\cal M}}

\newcommand{\mpi}{m_{\pi}}
\newcommand{\fpi}{f_{\pi}}
\newcommand{\fa}{f_{a}}
\newcommand{\Capi}{C_{a\pi}}

\definecolor{nicered}{rgb}{0.7,0.1,0.1}
\definecolor{nicegreen}{rgb}{0.1,0.5,0.1}
\definecolor{red}{rgb}{1.0, 0, 0}
\definecolor{niceblue}{rgb}{0,0,0.8}
\definecolor{red}{rgb}{1.0, 0, 0}
\hypersetup{colorlinks,bookmarksopen,bookmarksnumbered,
linkcolor=blus,pdfstartview=FitH,urlcolor=rossos,citecolor=verde}
\allowdisplaybreaks

\definecolor{rosso}{cmyk}{0,1,1,0.4}
\definecolor{rossos}{cmyk}{0,1,1,0.55}
\definecolor{rossoc}{cmyk}{0,1,1,0.2}
\definecolor{blu}{cmyk}{1,1,0,0.3}
\definecolor{blus}{cmyk}{1,1,0,0.6}
\definecolor{bluc}{cmyk}{1,1,0,0.1}
\definecolor{verde}{cmyk}{0.92,0,0.59,0.25}
\definecolor{verdec}{cmyk}{0.92,0,0.59,0.15}
\definecolor{verdes}{cmyk}{0.92,0,0.59,0.4}

\def\eq#1{{Eq.~(\ref{#1})}}
\def\eqs#1#2{{Eqs.~(\ref{#1})--(\ref{#2})}}
\def\fig#1{{Fig.~\ref{#1}}}

\def\sect#1{{Sect.~\ref{#1}}}

\def\app#1{{App.~\ref{#1}}}
\def\apps#1#2{{Apps.~\ref{#1}--\ref{#2}}}
\def\abs#1{\left| #1\right|}

\def\Tr{\mbox{Tr}}

\def\diag{\mbox{diag}}
\renewcommand{\d}{{\rm d}}
\renewcommand{\bar}{\overline}

\newcommand{\Ud}{U^\dagger}
\newcommand{\Chid}{\chi^\dagger}
\newcommand{\dmu}{\partial_\mu}

\newcommand{\dmuu}{\partial^\mu}

\newcommand{\Dnuu}{D^\nu}
\newcommand{\Dmu}{D_\mu}
\newcommand{\Dnu}{D_\nu}
\newcommand{\Dmuu}{D^\mu}
\newcommand{\thetaR}{\Theta ^ { a } _ { R }(x)}
\newcommand{\thetaL}{\Theta ^ { a } _ { L }(x)}
\newcommand{\fRd}{f _ { \mu \nu } ^ { R }}

\newcommand{\fLd}{f _ { \mu \nu } ^ { L }}

\newcommand{\pim}{\pi_-}
\newcommand{\pip}{\pi_+}
\newcommand{\piz}{\pi_0}
\newcommand{\pizb}{\pi_{0b}}

\newcommand{\beq}{\begin{equation}}
\newcommand{\eeq}{\end{equation}}
\newcommand{\bea}{\begin{eqnarray}}
\newcommand{\eea}{\end{eqnarray}}

\renewcommand{\[}{\left[}
\renewcommand{\]}{\right]}
\renewcommand{\(}{\left(}
\renewcommand{\)}{\right)}

\begin{document}

%\thispagestyle{empty}

%\begin{flushleft}
%\hskip 12cm DESY XX-XXX
%\end{flushleft}

\begin{center}  
%\vspace{3cm}
{\LARGE\bf\color{blus} 
$a \to \pi \pi \pi$ decay 
at 
%NLO 
next-to-leading order \\
in chiral perturbation theory 
%Axion-pion chiral Lagrangian at NLO: $a \to \pi \pi \pi$ decay
} \\
\vspace{1cm}
{\bf Luca Di Luzio$^{a,b}$, Gioacchino Piazza$^{c}$}\\[7mm]

{\it $^a$Dipartimento di Fisica e Astronomia `G.~Galilei', Universit\`a di Padova, Italy}\\[1mm]
{\it $^b$Istituto Nazionale Fisica Nucleare, Sezione di Padova, Italy}\\[1mm]
{\it $^c$Université Paris-Saclay, CNRS, IJCLab,  
91405, Orsay, France}\\[1mm]

\vspace{0.2cm}
\begin{quote}
We discuss the construction of the two-flavour 
axion-pion 
effective Lagrangian 
at the next-to-leading order (NLO) in chiral perturbation theory 
and  present,  
as a phenomenological application,  
%two phenomenological applications: 
%the scattering $a\pi \to \pi\pi$, 
%relevant for axion 
%thermalization in the early Universe, 
%and 
the calculation of 
the decay rate of a GeV-scale axion-like particle 
via the 
channel $a \to \pi\pi\pi$. 
Through the NLO calculation,   
we assess
the range of validity of the effective field theory and 
show that the chiral 
expansion breaks down 
just above the kinematic
threshold. 
%These results call for 
Alternative non-perturbative approaches 
are called for 
in order to extend 
the chiral description of axion-pion interactions. 
\end{quote}
\thispagestyle{empty}
\end{center}

\tableofcontents

\section{Introduction}
\label{sec:intro}

The main ingredient of the axion solution to the strong CP problem \cite{Peccei:1977hh,Peccei:1977ur,Weinberg:1977ma,Wilczek:1977pj} 
is the axion coupling to a pseudo-scalar gluon density, which sets model-independent experimental 
targets for the axion mass and 
couplings to photons, nucleons, pions and electrons. 
Since the axion is much lighter than the scale of chiral symmetry breaking $\Lambda_\chi \simeq 1$ GeV 
and it has the same quantum numbers of the neutral pion, chiral perturbation theory ($\chi$PT) provides a natural 
framework to systematically derive axion properties. 
In fact, those were obtained long time ago by 
using leading order (LO) $\chi$PT (or equivalently current algebra) 
in a series of renowned papers \cite{Weinberg:1977ma,Bardeen:1977bd,DiVecchia:1980yfw,Kaplan:1985dv,Srednicki:1985xd,Georgi:1986df}. 
The axion chiral potential and coupling to photons at the next-to-LO (NLO) in $\chi$PT were 
computed in Ref.~\cite{Spalinski:1988az} (see also \cite{Mao:2009sy}), 
but it is only more recently that the program of ``precision'' axion physics has restarted 
with Ref.~\cite{diCortona:2015ldu}, also motivated by the booming of the 
axion experimental program (see e.g.~\cite{Irastorza:2018dyq,Sikivie:2020zpn}). 
State of the art axion mass calculations 
are now obtained by employing next-to-NLO (NNLO) 
$\chi$PT \cite{Gorghetto:2018ocs} or, alternatively, 
via lattice QCD techniques \cite{Borsanyi:2016ksw}.  
The axion-nucleon interaction Lagrangian instead has been derived in 
heavy baryon $\chi$PT up to NNLO \cite{Vonk:2020zfh,Vonk:2021sit}. 
Also CP- and flavour-violating axion couplings have witnessed 
a resurgence of interest in the recent years, 
with new calculations based either on $\chi$PT or other 
non-perturbative approaches (see respectively Refs.~\cite{Bigazzi:2019hav,Bertolini:2020hjc,Okawa:2021fto,Dekens:2022gha} 
and \cite{MartinCamalich:2020dfe,Bauer:2021wjo,Guerrera:2021yss}). 

In this paper we focus on the axion-pion chiral Lagrangian at NLO. 
The latter was previously considered 
in Refs.~\cite{Spalinski:1988az,diCortona:2015ldu} in the context of the axion potential, 
hence limited to non-derivative axion interactions,
and more generally in Ref.~\cite{DiLuzio:2021vjd}, 
which included also derivative axion couplings.  
We here expand on the derivation of the NLO axion-pion chiral Lagrangian, 
by providing several details which were not presented in 
Ref.~\cite{DiLuzio:2021vjd}. 

The most interesting application of this formalism consists in the calculation of the 
$a\pi \to \pi\pi$ scattering, 
which provides the dominant channel for axion thermalization 
in the early Universe \cite{Chang:1993gm,Hannestad:2005df},  
when the axion decouples from the thermal bath 
at temperatures below that of QCD deconfinement 
$T_c \simeq 155$ MeV~\cite{Aoki:2006br,Borsanyi:2010bp,Bazavov:2011nk}. 
The highest attainable axion mass 
from cosmological constraints 
on thermally-produced axions 
is known as the axion hot dark matter bound. 
However, as shown in Ref.~\cite{DiLuzio:2021vjd}, 
the chiral expansion of the axion-pion thermalization rate breaks down well below $T_c$.
%this bound
%was 
%mainly
%extracted from a 
%temperature regime, $T \gtrsim 60$ MeV, 
%where the 
%chiral approach to axion-pion scattering breaks down. 
Lacking for the moment a way to extrapolate the validity of $\chi$PT, 
a practical solution was given in Refs.~\cite{DEramo:2021psx,DEramo:2021lgb} 
which proposed an interpolation of the thermalization rate starting from the high-temperature region above $T_c$. 
See Refs.~\cite{Caloni:2022uya,DEramo:2022nvb} 
for recent cosmological analyses adopting this latter approach. 

Another application of the axion-pion chiral Lagrangian, 
which is the main subject of this paper,
arises in the context of 
GeV-scale axion-like particles (ALPs) which dominantly decay hadronically 
as soon as the phase space for the channel $a \to \pi\pi\pi$ is kinematically open. 
For phenomenological studies related to this channel, see e.g.~Refs.~\cite{Aloni:2018vki,Kelly:2020dda,Cheng:2021kjg}.  
This process was computed at LO in $\chi$PT in Refs.~\cite{Bauer:2017ris,Bauer:2020jbp} 
and the chiral expansion was claimed to be valid up to ALP masses of 
%order $4\pi f_\pi \simeq 1.6$ GeV. 
few GeV.
However, by explicitly computing the NLO correction, we find that the effective field theory (EFT) breaks down 
much earlier, namely for ALP masses just above the kinematical threshold $m_a \gtrsim 3 m_\pi$. 
Hence, in practice, $\chi$PT never yields an accurate description for the process at hand. 

The paper is structured as follows: in \sect{sec:axionchiral} we discuss the construction of the 
axion-pion chiral Lagrangian, while the calculation of the 
%two processes  
%$a\pi \to \pi\pi$ and 
$a \to \pi\pi\pi$ decay up to NLO in $\chi$PT is outlined in \sect{sec:twoappl}. 
We conclude in \sect{sec:concl}, where we 
%highlight some general expectations for 
%the range of validity of the EFT and 
advocate for possible strategies in order to extend the validity of the chiral description. 
Further details on the NLO calculation are provided in \apps{sec:axialcurr}{sec:ALPdecayampl}.

\section{Axion-pion effective field theory}
\label{sec:axionchiral}

The construction of the LO axion-pion Lagrangian was originally discussed 
in Refs.~\cite{DiVecchia:1980yfw,Georgi:1986df}. 
We first recall its basic ingredients (see also \cite{Chang:1993gm,Bauer:2017ris,DiLuzio:2020wdo,Bauer:2020jbp}) 
in view of the extension at NLO, which was recently discussed in Ref.~\cite{DiLuzio:2021vjd}.  
We here complement the latter derivation
by providing several details which were omitted in Ref.~\cite{DiLuzio:2021vjd}. 
In particular, we will focus on the 2-flavour formulation, which is best suited for the 
application to be discussed in \sect{sec:twoappl}. 
This is justified a posteriori, 
because the presence of strange mesons 
as external states is kinematically suppressed 
up to the energy scale at which the chiral expansion breaks down. On the other hand, 
the generalization to the 3-flavour case is in principle straightforward. 
In the following we will generically indicate both the QCD axion and the ALP as ``axion'', 
specifying when needed which case we are considering.

\subsection{Axion-QCD effective Lagrangian}
\label{sec:axionquaekEFT}

The 2-flavour axion effective Lagrangian in terms of quarks and gluons reads 
\begin{align}\label{Leffaq}
    \mathcal{L}^{\rm QCD}_{a} &= \frac{1}{2}(\dmu a)^2 - \frac{1}{2} m_{a,0}^2 a^2 + \frac{\alpha_s}{8 \pi} \frac{a}{f_a} G \widetilde G 
- \bar q_R M_q q_L
  + \text{h.c.} \nonumber \\
&+\frac{\dmu a}{2 f_a}\bar q c_q^0 \gamma^\mu \gamma_5 q 
+ \frac{1}{4} g^0_{a\gamma} a F \tilde F 
\, , 
\end{align} 
where $q=(u,d)^T$, 
$M_q = \diag \, (m_u,m_d)$,  
$G \widetilde G \equiv \frac{1}{2} \epsilon^{\mu\nu\rho\sigma} G^A_{\mu\nu} G^A_{\rho\sigma}$ 
and $F \widetilde F \equiv \frac{1}{2} \epsilon^{\mu\nu\rho\sigma} F_{\mu\nu} F_{\rho\sigma}$,  
with $\epsilon^{0123} = -1$. 
For the QCD axion $m_{a,0}^2=0$, while $m_{a,0}^2 \neq 0$ for the ALP case.\footnote{In the ALP case there could be 
an extra term in the mass parameter of the 
type $- \frac{1}{2} m_{a,0}^2 (a-a_0)^2$ so that the ALP field does not relax in zero. 
Here, we do not specify the mechanism responsible for solving the strong CP problem in the ALP case 
%(either a tuning or a UV mechanism) 
and assume for simplicity $a_0 = 0$, since for $a_0 \neq 0$ the main observable computed in this paper 
is not affected at the leading order in $1/f_a$.} 
In the following, we will be especially interested in the case where $m_{a,0} \sim$ GeV, 
i.e.~much larger than the pure QCD axion mass contribution.
The couplings $c^0_{q} = \text{diag}(c^0_{u},c^0_{d})$ and $g^0_{a\gamma}$ are model-dependent. 
For instance, in the case of the QCD axion, $c^0_{u,d} = 0$ and $g^0_{a\gamma} = 0$ in the KSVZ model \cite{Kim:1979if,Shifman:1979if}, 
while $c^0_{u} = \frac{1}{3} \cos^2\beta$, $c^0_{d} = \frac{1}{3} \sin^2\beta$ and $g^0_{a\gamma} = \alpha/(2\pi f_a) 8/3$
in the DFSZ model \cite{Zhitnitsky:1980tq,Dine:1981rt} 
(with $\tan\beta = v_u / v_d$ the ratio between the 
vacuum expectation values 
of the two Higgs doublets present in the DFSZ model).

Upon an anomalous axial rotation of the quark doublet 
\begin{equation}
\label{eq:qtransf}
    q \rightarrow e^{i \gamma_5 \frac{a}{2 f_a} Q_a } q \, , 
\end{equation}
with $\Tr \, Q_a = 1$, 
the $aG\tilde G$ term in \eq{Leffaq} is shifted away, 
and the Lagrangian in \eq{Leffaq} becomes 
\begin{equation}\label{Leffaq2}
    \mathcal{L}^{\rm QCD}_{a}= \frac{1}{2}(\dmu a)^2 - \frac{1}{2} m_{a,0}^2 a^2 
-
(\bar q_R M_a q_L + \text{h.c.}) 
+\frac{\dmu a}{2 f_a}\bar q c_q \gamma^\mu \gamma_5 q 
+ \frac{1}{4} g_{a\gamma} a F \widetilde F \, , 
\end{equation} 
where we have redefined the parameters  as
\beq
\label{eq:Mqdressed}
 M_a = e^{-i \frac{a}{2 f_a} Q_a } M_q e^{-i \frac{a}{2 f_a} Q_a } \, , 
   \quad
   c_q = c_q^0 -Q_a  \, , \quad 
    g_{a\gamma} = g^0_{a\gamma} - \frac{3 \alpha}{2 \pi f_a} \Tr\, (Q_a Q_{\rm EM}^2) \, , 
\eeq
with 
$Q_{\rm EM} = \diag\,(2/3,-1/3)$.

\subsection{Axion-pion effective Lagrangian at LO}
\label{sec:axionLO}

At energies $\lesssim$ 1 GeV, 
the axion-QCD effective Lagrangian is replaced by the 
axion chiral Lagrangian, 
which at the LO reads (in the 2-flavour approximation, relevant for the observable studied in this paper)
\begin{align}\label{L_LO}
\mathcal{L}^{\chi({\rm LO})}_{a} &= \frac{1}{2}(\dmu a)^2 - \frac{1}{2} m_{a,0}^2 a^2 +\frac{f_\pi^2}{4} \Tr\left[(\Dmuu U)^\dagger \Dmu U + U\Chid_a +\chi_a \Ud  \right] \nonumber \\
& + \frac{\dmuu a}{2f_a}\Tr\left[c_q \sigma^a\right] J^{a}_{A,\, \mu} |_{\rm LO} \, ,
\end{align}
where $f_\pi = 92.21\ \rm MeV$, $\chi_a = 2 B_0 M_a$ 
(with $B_0$ denoting the quark condensate)  
and $\sigma ^a$ ($a=1,2,3$) the Pauli matrices. 
$U = e^{i\pi^a \sigma^a/f_\pi}$ is the pion Goldstone matrix, with  
\beq 
\pi^a \sigma^a = 
\begin{pmatrix}
\pi_0 & \sqrt{2} \pi_+ \\
\sqrt{2} \pi_- & - \pi_0 
\end{pmatrix} \, . 
\eeq
The pion axial current, $J^{a}_{A,\, \mu}$, reads at the LO (see \app{sec:axialcurr})

\begin{equation}\label{AxialCurr}
J^{a}_{A,\, \mu} |_{\rm LO} = \frac{i}{4}f_\pi^2 \Tr\left[\sigma^a \{U,(D_\mu U)^\dagger\}   \right] \, ,
\end{equation}
defined in terms of the covariant derivative $\Dmu U = \dmu U - i r_\mu U + i U \ell_\mu$, with 
$r_\mu=r_\mu^a \sigma^a/2$ and $l_\mu=l_\mu^a \sigma^a/2$ external fields which can be used to include electromagnetic or weak effects.
The matching of the derivative axion term in \eq{L_LO} with the corresponding one in \eq{Leffaq2} 
has been obtained by rewriting 
\begin{equation}
    \bar q_i [c_q]_{ij} \gamma^\mu \gamma_5 q_j = \frac{1}{2} \Big(\Tr\,[c_q] \underbrace{\bar q \gamma^\mu \gamma_5 q }_{\rm iso-singlet}+ \Tr\,[ c_q \sigma ^a] \underbrace{ \bar q \gamma^\mu \gamma_5\tfrac{\sigma^a}{2} q  }_{\rm iso-triplet} \Big)\, , 
\end{equation}
where we used the Fierz identity $\sigma^a_{ij} \sigma^a_{kl} = 2 ( \delta_{il}\delta_{jk}-\frac{1}{2}\delta_{ij}\delta_{kl})$. 
The iso-singlet current is associated to the heavy $\eta'$ and it can be neglected for our purposes, while the 
iso-triplet quark axial current is replaced with the pion axial current in \eq{AxialCurr}.  

In the following, we set $Q_a = M_q^{-1} / \Tr \, M_q^{-1}$, 
so that 
terms linear in $a$ 
(including $a$-$\pi^0$ mass mixing)
drop out from \eq{L_LO} 
and 
the only linear axion term 
arise from the derivative interaction 
with the 
pion axial current. 
Explicitly, the derivative axion coupling 
reads
\beq 
\Tr \[ c_q \sigma^a \] = 
\(\frac{m_u - m_d}{m_u + m_d} 
+ c^0_{u} - c^0_{d}\) \delta^{a3} \, .
\eeq 
Expanding the pion axial current 
$J_{A,\, \mu}^a |_{\rm LO} = f_\pi \dmu \pi^a -\frac{1}{f_\pi} \pi^2 \dmu \pi^a - \frac{3}{2 f_\pi} \pi^a \dmu \pi^2 
+ \ldots \ $, with $\pi= \sqrt{\piz \piz+2 \pip \pim}$,
the axion-pion derivative terms are given by 
\begin{align*}
\label{a-pi coupling}
         \frac{\dmuu a}{2f_a}&{\rm Tr}\left[c_q \sigma^a\right] J^{a}_{A,\, \mu} |_{\rm LO} \simeq
         - \frac{1}{2} \Bigg( \frac{m_d-m_u}{m_u+m_d} +c_d^0-c_u^0\Bigg) \frac{f_\pi}{f_a}\dmu a \dmuu \piz  \numberthis \\ 
         + &\frac{1}{3}\Bigg( \frac{m_d-m_u}{m_u+m_d} +c_d^0-c_u^0\Bigg)\frac{1}{f_a f_\pi} \dmu a \Big( 2 \partial^\mu \piz \pip \pim -\piz \partial^\mu \pip \pim -\piz \pip \partial^\mu \pim\Big)\, .
\end{align*}
The first operator introduces a kinetic mixing between 
the axion and the neutral pion, 
parametrized by the coefficient  
\begin{equation}
    \epsilon \equiv -\frac{1}{2}\frac{f_\pi}{f_a} \left(\frac{m_d-m_u}{m_d+m_u}+c_d^0-c_u^0\right) \, .
\end{equation}
At the quadratic level  
the $a$-$\pi^0$ Lagrangian reads 
\begin{equation}\label{Ldiag}
 \mathcal{L}_{a-\piz}^{\rm quad} = \frac{1}{2} \left(\dmu a \quad \dmu \piz\right) \mathcal{K}_{ \rm LO} \begin{pmatrix}
\dmuu a \\
\dmuu \piz 
\end{pmatrix} 
- \frac{1}{2} \left(a \quad \piz\right) \mathcal{M}^2_{\rm LO}
\begin{pmatrix}
 a \\
 \piz 
\end{pmatrix} \, ,
\end{equation}
with
\begin{equation}
    \mathcal{K}_{\rm LO}=\begin{pmatrix}
  1 & \epsilon \\
  \epsilon & 1
 \end{pmatrix}, \qquad
 \mathcal{M}^2_{\rm LO}=\begin{pmatrix}
  m_{a}^2 & 0 \\
  0  & m_{\pi}^2
 \end{pmatrix}
\end{equation}
and $m_a^2=m_{a,0}^2+m_{a, \text{QCD}}^2$, where  
\begin{equation}\label{ma}
    m_{a, \text{QCD}}^2 = \frac{m_u m_d}{(m_u+m_d)^2}\frac{\mpi^2 f_\pi^2}{f_a^2} 
    \simeq 5.7 \Bigg(\frac{10^{12} \ \text{GeV}}{f_a}\Bigg)\ {\rm \mu eV} \, , 
\end{equation}
is the QCD axion mass squared at the LO. 
The procedure in order to diagonalize the quadratic Lagrangian in \eq{Ldiag}
consists of three steps: 
$i)$ diagonalization of the kinetic term by an orthogonal transformation, 
$ii)$ re-scaling of the fields to have a canonical kinetic term and 
$iii)$ diagonalization of the mass matrix (rotated and re-scaled after steps $i)$ and $ii)$). 
The first orthogonal rotation 
\begin{equation}
  R_1=  \frac{1}{\sqrt{2}}\begin{pmatrix}
 -1 & 1 \\
 1 & 1 
\end{pmatrix}
\end{equation}
gives
\begin{equation}
R_1^T\mathcal{K}_{\rm LO} R_1 =\begin{pmatrix}
 1-\epsilon & 0 \\
 0 & 1+\epsilon 
\end{pmatrix} \, .
\end{equation}
Therefore the re-scaling is given by (fields need to be multiplied by the inverse of $W$)
\begin{equation}
  W =  \begin{pmatrix}
 \frac{1}{\sqrt{1-\epsilon}} & 0 \\
 0 & \frac{1}{\sqrt{1+\epsilon}} 
\end{pmatrix} \, .
\end{equation}
The action of $R_1$ and $W$ on the mass matrix puts it in the form 
\renewcommand\arraystretch{2.4}
\begin{equation}\label{mass}
\hat{\mathcal{M}}^{2}_{\rm LO}=W R_1^T\mathcal{M}_{\rm LO} ^2 R_1 W=\frac{1}{2} \begin{pmatrix}
 \dfrac{m_{a}^2 +m_{\pi}^2  }{ 1-\epsilon} & \dfrac{m_{\pi }^2-m_{a }^2}{\sqrt{1-\epsilon ^2}} \\
 \dfrac{m_{\pi }^2-m_{a }^2}{\sqrt{1-\epsilon ^2}} & \dfrac{m_{a }^2 +m_{\pi }^2}{1+\epsilon} \\
\end{pmatrix} \, ,
 \end{equation}
 whose eigenvalues are $\mpi^2$ and $m_a^2$ plus corrections of $O(\epsilon^2)$ for the pion and ALP masses, and $O(\epsilon^4)$ for the QCD axion mass (considering $m_a/\mpi\sim \mathcal{O}(\epsilon)$ in the QCD axion case). 
Denoting by $R_2$ the matrix that diagonalizes \eq{mass} as $R_2^T \hat{\mathcal{M}}^{2}_{\rm LO} R_2$, one obtains that the complete rotation 
that needs to be applied to the 
fields $(a,\piz)$ in order to fully diagonalize the quadratic Lagrangian in \eq{Ldiag} is given by 
  \begin{equation}
\mathcal{R}=\left(R_1 W R_2\right)^{-1} = \begin{pmatrix}
 1- \dfrac{    \epsilon^2  m_{a }^4}{2 (m_{a }^2-m_{\pi }^2)^2} & -\dfrac{ \epsilon  m_{\pi }^2}{m_{a }^2-m_{\pi }^2} \\
 \dfrac{ \epsilon  m_{a }^2  }{m_{a }^2-m_{\pi }^2} & 1+\dfrac{    \epsilon^2  m_{\pi }^4}{2(m_{a }^2-m_{\pi }^2)^2} \\
\end{pmatrix} \, .
\end{equation}
Neglecting $\mathcal{O}(\epsilon^2)$ terms in $\mathcal{R}^{-1}$,  
we finally obtain\footnote{Ref.~\cite{Bauer:2020jbp} provides a more general expression in a basis where $Q_a$ is only subject to the condition $\Tr \, Q_a = 1$, 
i.e.~without imposing $Q_a = M_q^{-1} / \Tr \, M_q^{-1}$.}
\begin{align}
 \label{fieldPhysLOa}
 a&= a_{\rm phys} +\dfrac{ \epsilon  m_{\pi }^2}{m_{a }^2-m_{\pi }^2} {\piz}_{\rm phys} \, , \\
  \label{fieldPhysLOpi}
 \piz&= {\piz}_{\rm phys}  -\dfrac{ \epsilon  m_{a }^2  }{m_{a }^2-m_{\pi }^2} a_{\rm phys} \, ,
 \end{align} 
where ($a_{\rm phys}$, ${\piz}_{\rm phys}$) denote fields with diagonal propagators.
In the following, we drop the subscript ``phys'' when working in the diagonal basis.

After the LO diagonalization procedure, the LO chiral Lagrangian containing the axion-pions interaction terms is given by 
(including the contribution due to \eq{fieldPhysLOpi} from the standard 4-pion Lagrangian)
\begin{align}\label{Lapi}
        &\mathcal{L}_{a\pi}^{\chi({\rm LO})}=\frac{C_{a\pi}}{2 f_a f_\pi  (m_{a }^2-m_{\pi }^2) } \Big\{ 
        (m_a^2-2m_\pi^2)\, \dmu a \Big( 2 \dmu \piz \pip \pim -\piz \dmu \pip \pim -\piz \pip \dmu\pim\Big) \nonumber \\
      &  + m_a^2\, a \Big(m_\pi^2 (\piz \pip\pim+\frac{1}{2}\piz^3)- 2 \piz \dmu \pip \dmuu \pim + \dmu \piz \left(\dmuu \pip \pim +\dmuu\pim \pip\right)\Big) 
      \Big\} \, , 
\end{align}
with 
\begin{equation}\label{Capi}
    C_{a\pi} = \frac{1}{3}\Bigg( \frac{m_d-m_u}{m_u+m_d} +c_d^0-c_u^0\Bigg)\, .
\end{equation}
The QCD axion case is recovered in the $m_a^2 \to 0$ limit.
%, since the squared axion mass is of $\mathcal{O}(\epsilon^2)$ and 
%hence it should be consistently neglected in \eq{fieldPhysLOpi}.
Note that the correction due to the kinetic mixing in \eq{fieldPhysLOpi} can be safely neglected in the QCD axion case 
since $m_a \ll m_\pi$.

\subsection{Axion-pion effective Lagrangian at NLO}
\label{sec:ALPNLO}

The axion-pion Lagrangian beyond LO requires two ingredients: 
the $\mathcal{O}(p^4)$ chiral Lagrangian with the axion-dressed coefficient $\chi_a = 2 B_0 M_a$ 
(cf.~\eq{eq:Mqdressed})
and the derivative axion interaction with the 
NLO pion axial current. Part of the material of this Section was previously presented in Ref.~\cite{DiLuzio:2021vjd}.
The 2-flavour chiral Lagrangian 
at $\mathcal{O}(p^4)$ 
can be expressed in various equivalent bases. Here we stick to the expression given by 
Gasser and Leutwyler \cite{Gasser:1983yg}, which in the standard trace notation reads \cite{Scherer:2002tk}
\begin{align} 
\label{eq:LchiNLO}
&\mathcal { L }^{\chi({\rm NLO})}_{a} =  \frac {\ell_ { 1 } } { 4 } \left\{ \operatorname { {\rm Tr} } \left[ D _ { \mu } U \left( D ^ { \mu } U \right) ^ { \dagger } \right] \right\} ^ { 2 } + \frac {\ell_ { 2 } } { 4 } \operatorname { {\rm Tr} } \left[ D _ { \mu } U \left( D _ { \nu } U \right) ^ { \dagger } \right] \operatorname { {\rm Tr} } \left[ D ^ { \mu } U \left( D ^ { \nu } U \right) ^ { \dagger } \right] \nonumber \\ & + \frac {\ell_ { 3 } } { 16 } \left[ \operatorname { {\rm Tr} } \left( \chi_a U ^ { \dagger } + U \chi_a ^ { \dagger } \right) \right] ^ { 2 } + \frac {\ell_ { 4 } } { 4 } \operatorname { {\rm Tr} } \left[ D _ { \mu } U \left( D ^ { \mu } \chi_a \right) ^ { \dagger } + D _ { \mu } \chi_a \left( D ^ { \mu } U \right) ^ { \dagger } \right] \nonumber \\ & +\ell_ { 5 } \left[ \operatorname { {\rm Tr} } \left( f _ { \mu \nu } ^ { R } U f _ { L } ^ { \mu \nu } U ^ { \dagger } \right) - \frac { 1 } { 2 } \operatorname { {\rm Tr} } \left( f _ { \mu \nu } ^ { L } f _ { L } ^ { \mu \nu } + f _ { \mu \nu } ^ { R } f _ { R } ^ { \mu \nu } \right) \right] \nonumber \\ & + i \frac { \ell_6 } { 2 } \operatorname { {\rm Tr} } \left[ f _ { \mu \nu } ^ { R } D ^ { \mu } U \left( D ^ { \nu } U \right) ^ { \dagger } + f _ { \mu \nu } ^ { L } \left( D ^ { \mu } U \right) ^ { \dagger } D ^ { \nu } U \right] \nonumber \\ & - \frac {\ell_ { 7 } } { 16 } \left[ \operatorname { {\rm Tr} } \left( \chi_a U ^ { \dagger } - U \chi_a ^ { \dagger } \right) \right] ^ { 2 } + \frac { h _ { 1 } + h _ { 3 } } { 4 } \operatorname { {\rm Tr} } \left( \chi_a \chi_a ^ { \dagger } \right) + \frac { h _ { 1 } - h _ { 3 } } { 16 } \left\{ \left[ \operatorname { {\rm Tr} } \left( \chi_a U ^ { \dagger } + U \chi_a ^ { \dagger } \right) \right] ^ { 2 } \right. \nonumber \\ &\left. + \left[ \operatorname { {\rm Tr} } \left( \chi_a U ^ { \dagger } - U \chi_a ^ { \dagger } \right) \right] ^ { 2 } - 2 \operatorname { {\rm Tr} } \left( \chi_a U ^ { \dagger } \chi_a U ^ { \dagger } + U \chi_a ^ { \dagger } U \chi_a ^ { \dagger } \right) \right\} - 2 h _ { 2 } \operatorname { {\rm Tr} } \left( f _ { \mu \nu } ^ { L } f _ { L } ^ { \mu \nu } + f _ { \mu \nu } ^ { R } f _ { R } ^ { \mu \nu } \right) \nonumber \\
& + \frac{\dmuu a}{2f_a}\Tr\left[c_q \sigma^a\right] J^{a}_{A,\, \mu} |_{\rm NLO} \, .
\end{align}
The low-energy constants $\ell_1, \ell_2, \ldots, \ell_7 $ are not fixed by chiral symmetry, 
but they need to be determined from experimental data or lattice QCD. 
The constants $h_1$, $h_2$, $h_3$ are coupled to pion-independent terms (see Eq.~(\ref{h-relation}) below). 
%Hence, in a theory without axions they would be of no physical relevance.  
The $f _ { \mu \nu } ^ { R,L }$ are the field strength tensors associated to the fields $r_\mu$ and $l_\mu$ 
appearing in the covariant derivative (see \cite{Gasser:1983yg} for details). 
Since we are only interested in processes involving an even number of bosons, we neglect here 
the  $\mathcal{O}(p^4)$ Wess-Zumino-Witten term \cite{Wess:1971yu, Witten:1983tw} which features intrinsic-parity-odd operators. 

The NLO chiral left (right) current is obtained by differentiating the NLO Lagrangian with respect to the external field $l_\mu$ ($r_\mu$). Taking the axial combination of the 
chiral currents 
(see \app{sec:axialcurr}) one obtains
\begin{align*}
\label{eq:Jax}
    J^a_{A,\, \mu} |_{\rm NLO} &=
    i \frac{\ell_1}{2} \Tr \[ \sigma^a \{\Dmu\Ud, U\}\] \Tr \[ \Dnu U \Dnuu\Ud \] \numberthis\\
    +i &\frac{\ell_2}{4} \Tr \[ \sigma^a \{\Dnuu\Ud, U \} \] \Tr \[ \Dmu U \Dnu\Ud + \Dnu U \Dmu \Ud \] \\
    -i &\frac{\ell_4}{8} \Tr \[ \sigma^a \{\Dmu U, \Chid_a\}- \sigma^a \{U, \Dmu \Chid_a \} +\sigma^a \{\Dmu \chi_a,\Ud\}-\sigma^a\{\chi_a,\Dmu \Ud\} \] \\
    + &\frac{\ell_6}{4} \Tr \[ \fRd [\sigma^a, \Dnuu U]\Ud  + \fRd U [\Dnuu \Ud , \sigma^a ]  + \fLd\Ud[\sigma^a, \Dnuu U] + \fLd [\Dnuu \Ud , \sigma^a] U \] \, . 
\end{align*}
Being interested only in axion-pion interactions, 
from now on we will set to zero the field strength tensors as well as the external currents.
Then the axion-pion Lagrangian up to NLO is given by the sum $\mathcal{L}^{\chi ({\rm LO})}_a+\mathcal{L}^{\chi ({\rm NLO})}_a$. 

Note that the NLO terms reintroduce 
a quadratic mixing of the axion field with the neutral pion. 
In \app{sec:1loopdiag} we explicitly repeat the diagonalization procedure at NLO, 
including as well one-loop terms from the LO chiral Lagrangian. 
In fact, the choice $Q_a=M_q^{-1}/{\rm Tr}\,M_q^{-1}$ 
allows us to eliminate only some of the mass mixing terms at NLO.
On the other hand, no axion-pion 
mixing arises from the term proportional to 
$h_1 - h_3$ in \eq{eq:LchiNLO}, since the latter does not depend on the pion field. 
This is readily seen by using the identity 
\begin{align}\label{h-relation}
&\left[{\rm Tr}\left(\chi_a\Ud+U\Chid_a\right)\right]^2+\left[{\rm Tr}\left(\chi_a\Ud-U\Chid_a\right)\right]^2-2 {\rm Tr}\left(\chi_a\Ud \chi_a \Ud + U\Chid_a U \Chid_a \right) \nonumber \\ 
&=\left[{\rm Tr}(\chi_a)\right]^2+\left[{\rm Tr}(\Chid_a)\right]^2 - \left[{\rm Tr}(\sigma^a\chi_a)\right]^2 - \left[{\rm Tr}(\sigma^a\Chid_a)\right]^2 \, .
\end{align}
The remaining axion-pion mass mixing is found to be
\begin{align}\label{Mix1}
    \mathcal{L}^{\chi ({\rm NLO})}_a \supset \mathit{a}\  \piz &\frac{ 3 C_{a \pi} m_\pi^4}{f_a \fpi (m_a^2-m_\pi^2)}
    \Big\{ - \ell_3 m_a^2 \nonumber \\
  & +\ell_7 \frac{1}{(m_d+m_u)^2} \[m_a^2(m_d^2+m_u^2-6m_d m_u) + 4 m_d m_u m_\pi^2 \] \Big\}\,.
\end{align}
 Considering instead derivative terms, 
at NLO the pion axial current gives rise to the following kinetic mixing term
 \begin{equation}\label{Mix3}
\frac{\dmuu a}{2f_a}{\rm Tr}\left[c_q \sigma^a\right] J^a_{A,\, \mu} |_{\rm NLO} \supset 
-\frac{3}{2} \ell_4 \frac{\mpi ^2}{f_a f_\pi}C_{a\pi}\dmu a \dmuu \piz \, .
\end{equation}
Besides those tree-level mixings, the axion and the neutral pion also mix through one-loop diagrams, 
generated by the LO terms in \eq{Lapi}.

\section{$a \to \pi\pi\pi$ decay at NLO}
\label{sec:twoappl}

As an application of the axion-pion chiral Lagrangian formalism at NLO we present 
here the calculation of the $a \to \pi \pi \pi$ decay rate, 
which shares some analogies with the case of $a \pi \to \pi \pi$ scattering discussed recently in Ref.~\cite{DiLuzio:2021vjd}. 
The decay $a \to \pi \pi \pi$ 
is one of the leading hadronic channels for GeV-scale ALPs, 
and it has been previously computed at the LO in Refs.~\cite{Bauer:2017ris,Bauer:2020jbp}. 
By means of the NLO correction we want to assess the convergence of the chiral expansion. 
%In this case we will show, through the NLO calculation, 
%that the $\chi$PT expansion breaks just above the kinematical threshold $m_a \gtrsim 3 m_\pi$. 
%The two calculations, which share similar technical issues, 

%\subsection{$a \rightarrow \pi\pi\pi$ decay}
%\label{sec:ALPto3pi}

The ALP decay rate in three pions is obtained by integrating the differential rate (see e.g.~Sect.~48 in \cite{Zyla:2020zbs})
\beq
\d \Gamma_{a\to3\pi}=\frac{1}{(2\pi)^3}\frac{1}{32m_a^3} \abs{\M_{a\to3\pi}}^2 \d u\, \d s\, , 
\eeq
where there are two possible decay channels: 
$a \to \piz \pip \pim $ and $a \to \piz \piz \piz$. 
In the following, we present the calculation of the ALP decay amplitudes and 
compare the LO to the NLO decay rate.

\subsection{LO amplitude} 
\label{sec:LOamplALP}

The LO amplitudes at $\mathcal{O}(1/f_a)$ are obtained from the interaction terms in \eq{Lapi} and are
found to be 
\begin{align}
\label{eq:MLOpip}
\M_{a\to\piz\pip\pim}^{\rm LO} &=\frac{3 C_{a\pi} m_{\pi }^2  \left(m_{\pi }^2-s\right)}{2 f_{\pi } f_a \left(m_a^2-m_{\pi }^2\right)} \, , \\
\label{eq:MLOpi0}
\M_{a\to \piz\piz\piz}^{\rm LO} &= -\frac{3 C_{a\pi} m_{\pi }^2 m_a^2}{2 f_{\pi } f_a \left(m_a^2-m_{\pi }^2\right)} 
\, ,
\end{align}
with the Mandelstam variables defined as  
\begin{equation}\label{Mand}
    \begin{split} 
s =& (p_1 + p_2)^2 = 2 p_1\cdot p_2 +\mpi^2 \, , \\ 
t =& (p_1- p_3)^2 = -2 p_1\cdot p_3 +\mpi^2 \, , \\ 
u =& (p_1- p_4)^2 = - 2 p_1\cdot p_4 +\mpi^2 \, . \\ 
 \end{split}
\end{equation}Note that the neutral pion channel (\eq{eq:MLOpi0}) 
is proportional to $m_a^2$, since it 
stems entirely from $a$-$\piz$ mixing.

\subsection{NLO amplitude} 
\label{sec:NLOamplALP}

\begin{figure}[tbp]
\centering
\includegraphics[width=16.5cm]{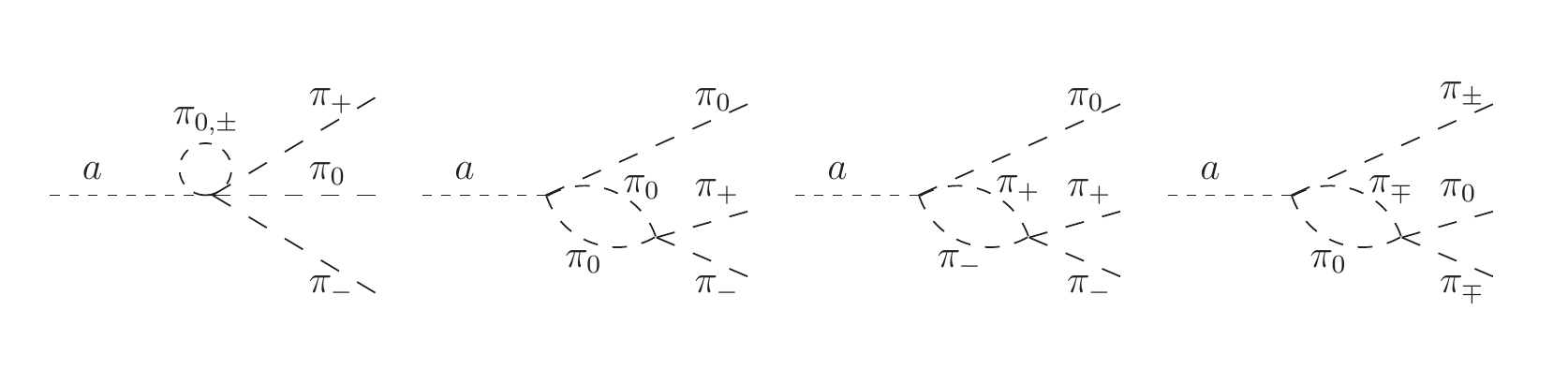}
\vspace{-1cm}
\caption{One-loop diagrams contributing to the ALP decay $a\to \piz \pip\pim$.}
\label{fig:ALPdecays}       
\end{figure}

%The calculation will be carried out 
%by retaining the axion-pion mixing at NLO explicitly, i.e.~through the 
%Lehmann-Symanzik-Zimmermann (LSZ) formalism \cite{Lehmann:1954rq}. 
%In \app{sec:1loopdiag} we provide instead an alternative calculation 
%%(only for the case of $a \pi \to \pi \pi$ scattering) 
%in which the axion-pion mixing is first removed via a direct NLO diagonalization of the 
%axion-pion propagator. 

To compute the ALP decay into three pions at NLO 
we employ the Lehmann-Symanzik-Zimmermann (LSZ) formalism \cite{Lehmann:1954rq}, according to which the amplitude is given by

\beq \label{Eq:AmpLSZALP}
\mathcal{M}_{a\to \piz \pi_i \pi_j }=\frac{1}{\sqrt{Z_a Z_\pi^3}}\Pi_{\alpha=1}^{4} \lim_{p_\alpha^2 \to m_\alpha^2} \left(p_\alpha^2-m_\alpha^2\right) \times G_{a\to \piz \pi_i \pi_j } (p_1,p_2,p_3,p_4) \, ,
\eeq
where the index $\alpha$ runs over the external particles, $(i,j)=(+,-)$ or $(0,0)$, and
$Z_a$ ($Z_\pi$) is the wave-function renormalization 
of the axion (pion) field defined via the residue of the 2-point Green's functions 
\beq
G_{\alpha\alpha}(p_\alpha^2 \simeq m_\alpha^2) = \frac{i Z_\alpha}{p_\alpha^2-m_\alpha^2} \, ,
\eeq
while the full 4-point Green's function 
is given by 
\beq
 \label{GreenF}
G_{a\to \piz \pi_i \pi_j}  =\sum_{k=a,\piz}\G_{k \to \piz \pi_i \pi_j }\times 
G_{a k}(m^2_{a} = 0)
G_{\pi_i \pi_i}(m^2_{\pi}) G_{\pi_j \pi_j}(m^2_{\pi})  G_{\piz \piz }(m^2_{\pi}) \, .
\eeq
The first term is the amputated 4-point function, multiplied by the 2-point 
functions of the external legs with the axion mass set to zero. 
We work in a basis where the $a$-$\pi_0$ mixing has been diagonalized at the lowest-order,
$\mathcal{O}(p^2)$, via \eqs{fieldPhysLOa}{fieldPhysLOpi} and the 
remaining mixing, of $\mathcal{O}(p^4)$, is retained explicitly.

%\footnote{\Red{In \app{sec:1loopdiag} we provide an alternative calculation %(only for the case of $a \pi \to \pi \pi$ scattering) in which the axion-pion mixing is first removed via a direct NLO diagonalization of the axion-pion propagator.} } 

Working with LO diagonal propagators, the 2-point amplitude 
for the $a\text{-}\piz$ system reads 

\beq 
\P_{ij} = \diag(p^2, p^2-m^2_\pi) - \Sigma_{ij} \, , 
\eeq
where $\Sigma_{ij}$ encodes NLO corrections including mixings. 
The 2-point Green's function
%, neglecting $\mathcal{O}(1/f_a^2)$,
is hence 
\beq
\label{2points}
G_{ij} =(-i \P)^{-1}_{ij}
=i \begin{pmatrix} 
\frac{1}{p^2} &\frac{\Sigma_{a\pi}}{p^2\(p^2-m_\pi^2-\Sigma_{\pi\pi}\)}\\
\frac{\Sigma_{a\pi}}{p^2\(p^2-m_\pi^2-\Sigma_{\pi\pi}\)} & \frac{1}{p^2-m_\pi^2-\Sigma_{\pi\pi}}
\end{pmatrix} \, . 
\eeq
Expanding the diagonal terms around the physical masses we get (see \eq{Zpi})
\beq
Z_a =1 \qquad Z_\pi= 1+ \Sigma'_{\pi\pi}(\mpi^2) \, , 
\eeq
with primes indicating derivatives with respect to $p^2$. 
Then, by plugging \eq{GreenF} and (\ref{2points}) into the LSZ formula for the scattering amplitude 
and neglecting $\mathcal{O}(1/f_a^2)$ terms,
%n \eq{Eq:AmpLSZ} 
%with ALP-pion propagators diagonalized at the LO. 
%Considering theALP-pion mixings at the NLO, 
we obtain the ALP-decay amplitudes which are given by 
\beq
\label{eq:DecayAmp}
\M_{a\to \piz \pi_i \pi_j }=\(1+\frac{3}{2} \Sigma'_{\pi\pi}\) \G_{a\to\piz\pi_i\pi_j}^{\rm LO} 
+\frac{\Sigma_{a\pi}(p^2=m_a^2)}{m_a^2-\mpi^2}\G_{\piz \to \piz\pi_i\pi_j}^{\rm LO}+ \G_{a\to\piz\pi_i\pi_j}^{\rm NLO} \, .
\eeq
Defining the invariant mass of the two-pions systems $\pi_\alpha$-$\pi_\beta$ as $(p_{\pi_\alpha}+p_{\pi_\beta})^2=m_{\alpha \beta}^2\equiv s,t,u$ 
with, respectively, $(\alpha, \beta)=(+,-),(0,+),(0,-)$, we obtain 
\begin{align}
\G_{a\to\piz\pip\pim}^{\rm LO} =&\frac{3 C_{a\pi} m_{\pi }^2  \left(m_{\pi }^2-s\right)}{2 f_{\pi } f_a \left(m_a^2-m_{\pi }^2\right)}  \, , \\
%+\frac{3 C_{a\pi} \mpi^4 (m_a^2 (\bar{\ell_3}+2\bar{\ell_4})-2 \bar{\ell_4} \mpi^2)  \left(\mpi^2-s\right)}{64 \pi ^2 f_a f_\pi^3 \left(m_a^2-\mpi^2\right) } \, , \\
\G_{\piz\to\piz\pip\pim}^{\rm LO}=& \frac{m_a^2+2 m_{\pi }^2-3 s}{3 f_{\pi }^2} \, , \\
\G_{a\to \piz\piz\piz}^{\rm LO} = &-\frac{3 m_{\pi }^2 m_a^2 C_{a\pi}}{2 f_{\pi } f_a \left(m_a^2-m_{\pi }^2\right)} \, , \\
% -\frac{3 C_{a\pi}m_a^2 m_{\pi}^4 (\bar{\ell_3}+2 \bar{\ell_4} )}{64 \pi ^2 f_a f_\pi^3 \left(m_a^2-m_{\pi}^2\right) } \, , \\
\G_{\piz \to\piz\piz\piz}^{\rm LO} =&  -\frac{m_{\pi }^2}{f_{\pi }^2}  \, ,\\
\Sigma'_{\pi\pi}=& \frac{2 I}{3 f_\pi^2} \, , \\
 \Sigma_{a\pi} (p^2)  = &\frac{C_{a\pi} }{f_\pi f_a}\Bigg(\frac{3 \ell_3 m_{\pi }^4 m_a^2}{ m_a^2-m_{\pi }^2 }+\frac{3 \ell_4 m_{\pi }^2 p^2}{2 }-\frac{I \left(4 p^2 \left(m_a^2-2 m_{\pi }^2\right)+m_{\pi }^2 m_a^2\right)}{4  \left(m_a^2-m_{\pi }^2\right)}\nonumber \\
 &
-\frac{
 3\ell_7 m_{\pi }^4  \left(m_a^2 \left(m_d^2 +m_u^2 -6 m_d m_u \right) +4 m_{\pi }^2 m_d m_u\right)
 }{
  \left(m_a^2-m_{\pi }^2\right) \left(m_d+m_u\right){}^2    } \Bigg) \, ,
\end{align}
with $I$ defined in \eq{Int}. The one-loop diagrams entering the Green's function $\G_{a\to\piz\pip\pim}^{\rm NLO} $ are shown in \fig{fig:ALPdecays},  and the full NLO decay amplitudes are reported in \eqs{eq:amp1}{eq:amp2}.

To carry out the renormalization procedure in dimensional regularization we shift the LECs as in \eq{eq:lbar} and we fix
$\gamma_1=1/3$, $\gamma_2=2/3$, $\gamma_3=-1/2$, 
$\gamma_4=2$ and $\gamma_7=0$,
consistently with the values found in the literature for the standard chiral theory \cite{Gasser:1983yg}.

\subsubsection{ALP decay rate: LO vs.~NLO }

\begin{figure}[tbp]
\centering
\includegraphics[width=7.5cm]{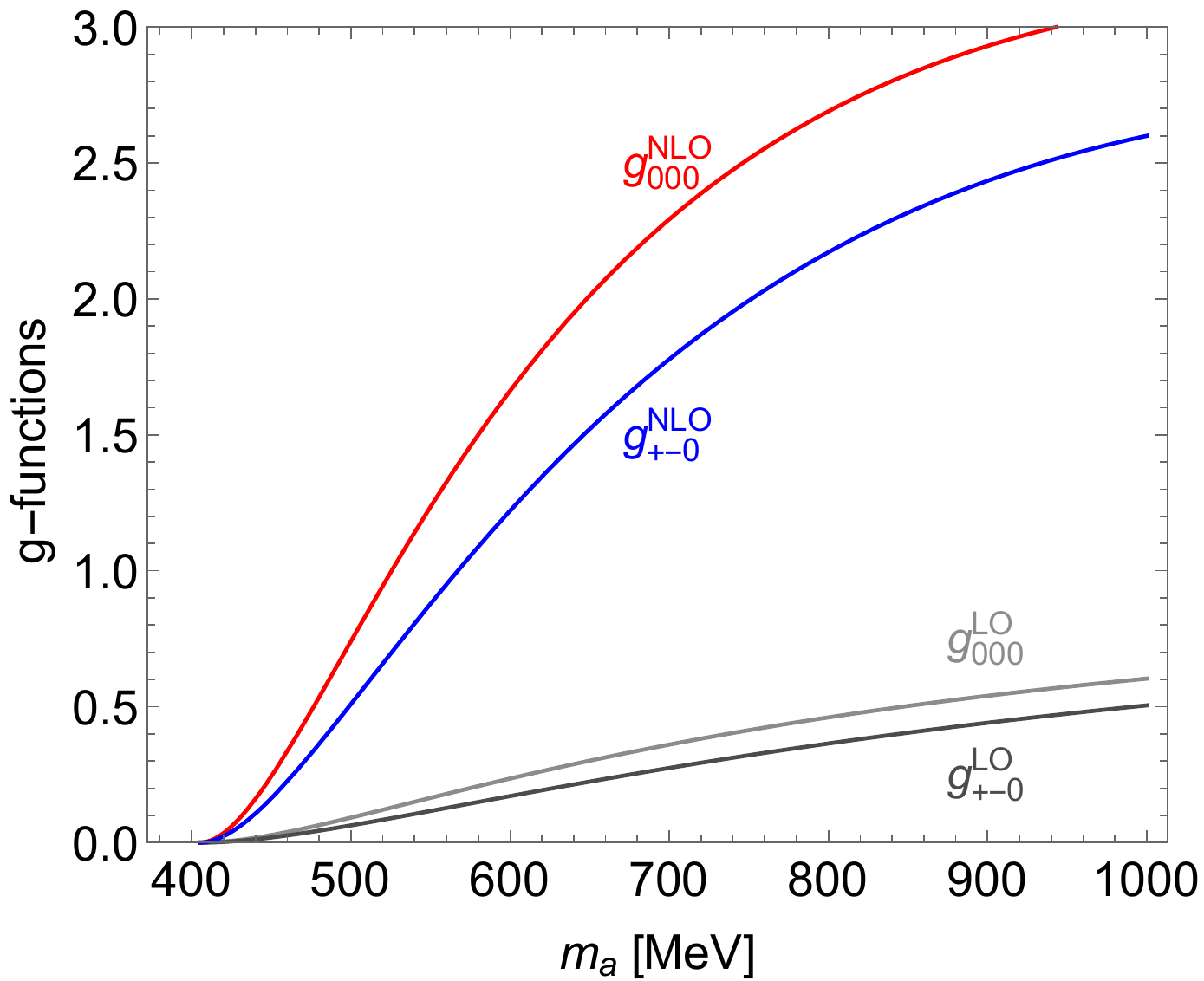}
\ \ \ \
\includegraphics[width=7.65cm]{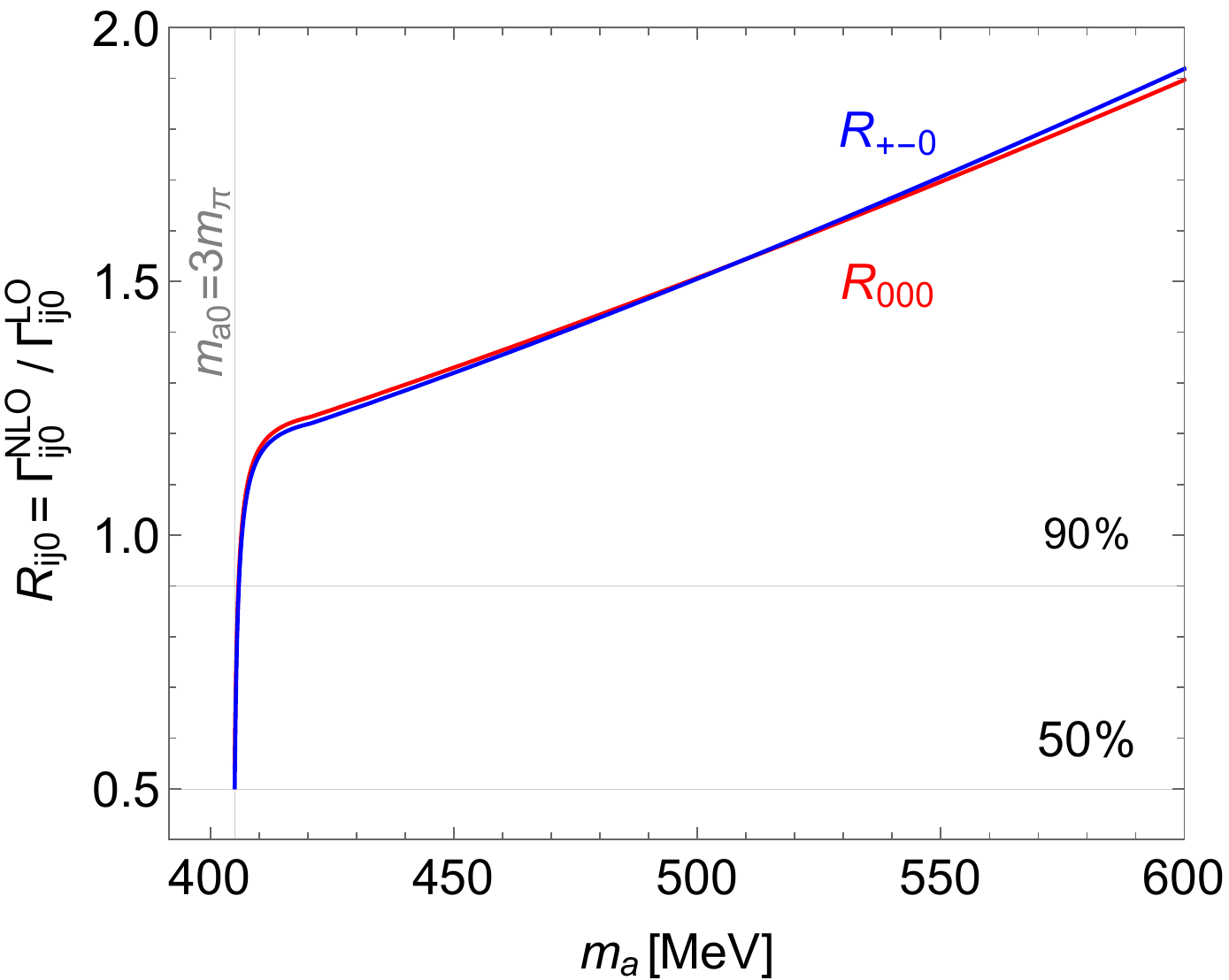}
\caption{Left panel: 
Numerical profile of $g^{\text{NLO}}_{000}$ and $g^{\text{NLO}}_{+-0}$, 
in red and blue respectively, 
compared to their LO counterparts in light and dark grey. 
Right panel: Ratio of the NLO to LO rates for the two possible decay channels. The vertical grey line indicates the 
kinematical threshold for the $a \to \pi\pi\pi$ decay, with $m_\pi=137$ MeV corresponding to the average neutral/charged pion
mass (at leading order in the isospin breaking).}
\label{fig:gALP}
\end{figure}

At LO we reproduce the decay rates given in Refs.~\cite{Bauer:2017ris,Bauer:2020jbp}, that in our notation read
\beq
\Gamma_{a\to\pi_i \pi_j \pi_0} ^{\text{LO}}= \frac{3 \Capi^2}{4096 \pi^3} \,\frac{m_a \mpi^4}{\fpi^2 \fa^2}\, g^{\text{LO}}_{ij0}(m_a) \, ,
\eeq
 with the numerical functions $g^{\text{LO}}_{ij0}(m_a)$ shown in the left panel of \fig{fig:gALP}. 
 Note that the $g^{\text{LO}}_{000}(m_a)$ function includes the symmetry factor $1/6$.

At NLO we only need to consider the interference between LO and NLO amplitudes, 
since NLO$^2$ terms are formally of higher order. 
For the numerical evaluation we used the 
central values of the
LECs 
$\overline{\ell_1} = -0.36(59)$ \cite{Colangelo:2001df}, 
$\overline{\ell_2} = 4.31(11)$ \cite{Colangelo:2001df},
$\overline{\ell_3} = 3.53(26)$ \cite{Aoki:2019cca},
$\overline{\ell_4} = 4.73(10)$ \cite{Aoki:2019cca}
and 
$\ell_7 = 2.5(1.4) \times 10^{-3}$ \cite{Frezzotti:2021ahg}, 
$m_u / m_d = 0.50(2)$ \cite{Aoki:2019cca}, 
$f_\pi = 92.1(8)$ MeV \cite{Zyla:2020zbs} and $m_\pi = 137$ MeV 
(corresponding to the average neutral/charged pion mass). 
Then the LO+NLO rates can be written as
\beq
\label{eq:ALPdecayLOpNLO}
\Gamma_{a\to\pi_i \pi_j \pi_0} ^{\text{LO+NLO}} 
= \frac{3 \Capi^2}{4096 \pi^3} \,\frac{m_a \mpi^4}{\fpi^2 \fa^2}\, 
\[ g^{\text{LO}}_{ij0}(m_a) + \frac{1}{16 \pi^2} \frac{m_a^2}{f_\pi^2} g^{\text{NLO}}_{ij0}(m_a) \] \, ,
\eeq
where the NLO functions $g^{\text{LO}}_{ij0}$ are obtained by numerically integrating 
the NLO amplitudes in \eqs{eq:amp1}{eq:amp2}.   
Their profile is shown in the left panel of \fig{fig:gALP}, 
for comparison with the LO counterparts. 
Although the expansion parameter in \eq{eq:ALPdecayLOpNLO} is formally 
written as $(m_a / 4\pi f_\pi)^2$, the actual calculation of the NLO rate 
shows (cf.~right panel of \fig{fig:gALP}) that the NLO correction becomes of the same order of the LO result 
already for ALP masses just above the 
kinematical threshold $m_a \gtrsim 3 m_\pi$. 
This is reflected by a somewhat larger value of the 
NLO $g$-functions compared to the LO ones, as shown in \fig{fig:gALP}.

Thus we conclude the $\chi$PT description of the $a \to \pi\pi\pi$ decay rate 
breaks down for ALP masses much smaller than $4\pi f_\pi \simeq 1.2$ GeV. 
%\cite{Bauer:2020jbp}. 
This earlier breakdown of $\chi$PT is also found 
in SM processes that are similar to the ALP decay into pions considered here,    
as e.g.~$\eta \to \pi \pi \pi$ (see e.g.~\cite{Bijnens:2007pr,Bijnens:2002qy}).
For instance, the NLO (NNLO) rate for $\eta \to \pi \pi \pi$ was found to be a factor $\approx 2.7$ $(4.5)$ larger than the LO one \cite{Bijnens:2007pr}. 

\section{Conclusions}
\label{sec:concl}

In this paper we have discussed the formulation of the axion-pion Lagrangian 
at the NLO in $\chi$PT and considered 
as an application of phenomenological relevance 
the ALP decay $a \to \pi \pi \pi$, which is one of the main hadronic channels for GeV-scale ALPs. 
Through the inclusion of the NLO correction, we have estimated
the range of applicability of the chiral expansion and found that the chiral EFT fails for ALP masses 
just above the kinematical threshold of $3m_\pi$ (cf.~right panel in \fig{fig:gALP}). 
This result shows an earlier  
breakdown of the chiral EFT compared to  
naive expectations based on previous LO calculations, 
see 
%respectively Refs.~\cite{Chang:1993gm,Hannestad:2005df} for $a \pi \to \pi \pi$ and 
Refs.~\cite{Bauer:2017ris,Bauer:2020jbp}.  
%for $a \to \pi \pi \pi$.  
We conclude that the range of applicability of the axion-pion chiral Lagrangian 
is rather limited for the problem at hand 
%two relevant processes discussed above and 
(similar conclusions were achieved for the case of $a \pi \to \pi \pi$ scattering in Ref.~\cite{DiLuzio:2021vjd})
and hence alternative non-perturbative approaches 
(based either on dispersion relations or lattice QCD techniques) 
are called for in order to extend the validity of the chiral description.

\begin{small}

\section*{Acknowledgments} 

We thank Guido Martinelli for many enlightening discussions on the subjects of this paper, 
%as well as 
and Ennio Salvioni for useful comments on the manuscript. We also thank Claudio Toni for spotting a mistake in the previous version of the paper.
The work of L.D.L.~and G.P.~has received funding from the European Union's Horizon 2020 research and innovation programme under the Marie Sk\l{}odowska-Curie grant agreement No 860881-HIDDEN.

\appendix

\section{Pion axial current}
\label{sec:axialcurr}

In this Appendix we provide the derivation of the pion axial current at the NLO. 
The currents associated to the Left and Right chiral rotations 
\begin{equation}
L = \exp \left( - i \Theta_L^a \frac { \sigma^a } { 2 } \right) \, , \qquad
R= \exp \left( - i \Theta_R^a \frac { \sigma^a } { 2 } \right) \, ,
\end{equation}
acting on the Goldstone matrix as $U \to R U L^\dag$, are easily computed promoting the global symmetries to local ones, and computing the variation $\delta \mathcal{L}$ of the Lagrangian under the given transformation.
From Noether's theorem, the Left and Right currents are given by 
\begin{equation}\label{J}
J_{L,R}^{\mu,a} = -\frac{\partial \delta \mathcal{L}}{\partial \partial_{\mu}\Theta ^ { a } _ { L,R }(x)}\, .
\end{equation}
Let us consider the LO chiral Lagrangian
\begin{equation}\label{Lchi}
\mathcal{L}_\chi = \frac{f_\pi^2}{4} {\rm Tr}\left[\dmuu U^\dagger \dmu U + U\Chid +\chi \Ud  \right] \, .
\end{equation}
To compute e.g.~the Right current, we set $\thetaL=0$ and perform an infinitesimal Right transformation 
\beq
U \rightarrow \left(1-i \thetaR \frac { \sigma^a } { 2 } \right)U \, . 
\label{lm}
\eeq
The variation of $\mathcal{L}_\chi$ is 
\begin{equation}
    \delta \mathcal{L}_\chi = \frac{i}{4} f_\pi^2 \dmu \thetaR {\rm Tr}\left[\dmuu U \Ud \sigma^a \right] \, ,
\end{equation}
and therefore $ J_{R}^{\mu,a}$ is given by
\begin{equation}
    J_{R}^{\mu,a} = -\frac{i}{4} f_\pi^2 {\rm Tr}\left[\dmuu U \Ud \sigma^a \right] \, .
\end{equation}
With an analogous procedure one obtains
\begin{equation}
    J_{L}^{\mu,a} = -\frac{i}{4} f_\pi^2 {\rm Tr}\left[\dmuu \Ud U \sigma^a \right] \, .
\end{equation}
The $R-L$ combination of these two currents provides the pion axial current at LO
\begin{equation}
 J_{A}^{\mu,a} = \frac{i}{4}f_\pi^2 {\rm Tr}\left[\sigma^a \{U,\dmuu U^\dagger\}   \right] \, .
\end{equation}
The procedure can be repeated at the NLO by employing the shift of the $\mathcal{O}(p^4)$ chiral Lagrangian in \eq{eq:LchiNLO}, 
which yields 
\begin{align*}
J_{R\ (\rm NLO)}^{\mu,a} = 
+i &\frac{\ell_1}{2} \Tr\[ \sigma^a U\Dmuu\Ud \] \Tr\[ \Dnu U \Dnuu\Ud \] \numberthis\\
+i &\frac{\ell_2}{4} \Tr\[ \sigma^a U\Dnuu\Ud \] \Tr\[ \Dmuu U \Dnu\Ud + \Dnu U \Dmuu \Ud \] \\
+i &\frac{\ell_4}{8} \Tr\[ \sigma^a \Dmuu U \Chid - \sigma^a U \Dmuu \Chid +\sigma^a \Dmuu \chi \Ud -\sigma^a \chi \Dmuu U^\dagger \] \\
 +&\frac{\ell_6}{4} \Tr\[ \fRd \left( \sigma^a \Dnuu U U^\dagger + U \Dnuu U^\dagger \sigma^a\right)  + \fLd \left(\Ud \sigma^a \Dnuu U + \Dnuu \Ud \sigma^a U \right) \] \, ,
\end{align*}
and
\begin{align*}
J_{L \ (\rm NLO)}^{\mu,a} = 
-i &\frac{\ell_1}{2} \Tr\[ \sigma^a \Dmuu\Ud U \] \Tr\[ \Dnu U \Dnuu\Ud \] \numberthis\\
-i &\frac{\ell_2}{4} \Tr\[ \sigma^a \Dnuu\Ud U \] \Tr\[ \Dnu U \Dmu\Ud + \Dmu U \Dnu \Ud \] \\
-i &\frac{\ell_4}{8} \Tr\[ \sigma^a \Chid \Dmuu U - \sigma^a \Dmuu \Chid U + \sigma^a \Ud \Dmuu \chi -\sigma^a \Dmuu \Ud \chi \] \\
+ &\frac{\ell_6}{4} \Tr\[ \fRd \left( \Dnuu U \sigma^a \Ud + U \sigma^a \Dnuu \Ud \right)  + \fLd \left(\Ud \Dnuu U \sigma^a+ \sigma^a \Dnuu \Ud U \right) \] \, .
\end{align*}
Combining the left and right currents we obtain the axial current in \eq{eq:Jax}.

\section{Axion-pion mixing at NLO}
\label{sec:1loopdiag}

We explicitly perform here the NLO diagonalization of the axion and neutral pion propagators.
%focussing for simplicity on the case of the QCD axion, for which $m^2_a \ll m^2_\pi$. 
%\subsection{NLO diagonalization}\label{NLOdiag}
The axion-neutral pion Lagrangian up to  
 order $1/f_a$ 
 %(hence neglecting $m_a^2$ terms) 
 is given by
\begin{equation}
    \mathcal{L}_{\text{$a$-$\piz$}} = \frac{1}{2} \left(\dmu a\right)^2+\frac{1}{2} \left(\dmu \pizb\right)^2 - \frac{1}{2} m_{a}^2 a^2- \frac{1}{2} m_{\pi}^2 \pizb^2 + \mathcal{L}_{\rm int}
\end{equation}
where the subscript $b$ stands for bare fields\footnote{We dropped the $b$ subscript for the axion field, 
since quantum corrections of $\mathcal{O}(1/f_a^2)$ are systematically neglected.}
and the interaction Lagrangian reads explicitly 
\begin{align}\label{Interaction}
    \mathcal{L}_{\rm int} &=  - a \pizb \ \ell_7\frac{4 m_d m_u m_{\pi }^4 \left(m_d-m_u\right)}{f_a f_\pi \left(m_d+m_u\right){}^3} 
    -\frac{3 C_{a\pi}}{2 f_a}\dmu a \dmuu \pizb \left(f_\pi+\ell_4\frac{\mpi^2}{f_\pi}\right) \nonumber \\ 
    +&\frac{2 C_{a\pi}}{f_a f_\pi }\dmu a \ \dmuu \pizb\ \pip\ \pim
    +\frac{1}{24 f_\pi^2}\mpi^2 \pizb^4-\frac{1}{3f_\pi^2}\pizb^2\dmu\pip\dmuu \pim \nonumber \\
    +&\frac{1}{6f_\pi^2}\mpi^2\pip\pim\pizb^2-\frac{1}{3f_\pi^2}\pip\pim\dmu\pizb\dmuu\pizb - \frac{\ell_3 \mpi^4 \pizb^2}{f_\pi^2} +\frac{\ell_7\pizb^2(m_d-m_u)^2\mpi^4}{f_\pi^2(m_u+m_d)^2}\, . 
\end{align}
Note that $\mathcal{L}_{\rm int}$ 
contains all the terms which contribute to the two-point functions of the neutral scalar fields, i.e.~LO tree-level mixings, LO terms giving the one-loop corrections and NLO terms. 
The latter provide the counterterms needed to reabsorb the loop divergences. 

We next define the renormalization conditions. 
Firstly, it is important to note that, since the divergences come from loops of $\mathcal{L}^{\chi({\rm LO})}_{a}$, it is sufficient to extract the counterterms from  $\ell_3$, $\ell_4$ and $\ell_7$.
Hence, $m_\pi$ and $f_\pi$ are the \textit{physical} pion mass and decay constant at LO.
%Firstly, it is important to note that in 
%$\chi$PT
%the pion mass $\mpi$ as well as the pion decay constant $f_\pi$ do not get renormalized. 
%This can be easily understood by observing that, 
%at NLO, $\mpi$ and $f_\pi$ are corrected by terms proportional to the bare low-energy constants $\ell_3$, $\ell_4$ and $\ell_7$ (see \eqs {mpiNLO}{fpiNlO}). Therefore, since the divergences come from loops of $\mathcal{L}^{\chi({\rm LO})}_{a}$, it is sufficient to extract the counterterms from  $\ell_3$, $\ell_4$ and $\ell_7$.
%Hence, $\mpi$ and $f_\pi$ are the \textit{physical} pion mass and decay constant at LO. 
Let us now denote by $-i\Sigma_{ij}(p^2)$ (with $i,j= a, \piz$) the 1-particle-irreducible (1PI) self-energy correction. 
The net effect of this correction is encoded in the effective Lagrangian 
\begin{equation}\label{Leffapiz}
\begin{split}
    \mathcal{L}_{\text{$a$-$\piz$}} ^{\rm eff} =& \frac{1}{2} a\left( p^2-m_a^2\right)a+\frac{1}{2} \piz\left( p^2 - m_{\pi}^2 + (p^2-\mpi^2)\delta Z_\pi - \Sigma_{\pi\pi}(p^2)\right) \piz \\ -& a \Sigma_{a\pi}(p^2)\left(1+\frac{1}{2}\delta Z_\pi\right)\piz \, , 
\end{split}
\end{equation}
where we employed the pion wave-function renormalization,  
$\pizb \rightarrow (1+\frac{1}{2}\delta Z_\pi)\piz$, 
defined as $\delta Z_{\pi}=\partial \Sigma_{\pi\pi}(p^2)/\partial p^2$. 
The one-loop self-energies $\Sigma_{ij}(p^2)$ can be computed 
from the interaction Lagrangian in \eq{Interaction}. Defining
\begin{equation}\label{Int}
   I=\frac{\mpi^2}{16\pi^2}\left[R+\log \left(\frac{\mpi^2}{\mu^2}\right)\right],\\    
\end{equation}
with $R= \frac{2}{d-4} - \log (4\pi)+\gamma_E-1$, and using dimensional regularization we find
\begin{align}
      \Sigma_{\pi\pi}(p^2) & = I\frac{1}{6f_\pi^2}[4p^2-\mpi^2]+\frac{2\ell_3\mpi^4}{f_\pi^2}-\frac{2\ell_7(m_d-m_u)^2\mpi^4}{f_\pi^2(m_u+m_d)^2}\label{Sigmap}\, ,\\
      \Sigma_{a\pi} (p^2) & = 3 C_{a\pi} p^2\left[\frac{f_\pi}{2f_a}+ \frac{\ell_4 \mpi^2}{2 f_a f_\pi}- \frac{2 I}{3 f_a f_\pi} \right]
    +\frac{4 \ell_7 (m_d-m_u)m_u m_d \mpi^4 }{f_a f_\pi (m_u+m_d)^3}\, ,\label{Sigmaa}
\end{align}
from which we get
\begin{equation}\label{Zpi}
    \delta Z_\pi = \frac{2 I}{3 f_\pi^2}\, .
\end{equation}
Therefore, we define the scale-independent parameters $\overline{\ell_i}$ and $\overline{h_i}$ in such a way 
that the 
$R+ \log(\mpi^2 /\mu^2)$ factor is subtracted \cite{Gasser:1983yg}
\begin{equation}\label{eq:lbar}
   \begin{split}
        \ell_i=&\frac{\gamma_i}{32\pi^2}\left[\overline{\ell_i}+R+\log \left(\frac{\mpi^2}{\mu^2}\right) \right]\, ,\\
        h_i=&\frac{\delta_i}{32\pi^2}\left[\overline{h}_i+R+\log \left(\frac{\mpi^2}{\mu^2}\right) \right]\, .
   \end{split}
\end{equation}
Plugging these definitions in \eqs{Sigmap}{Sigmaa} and substituting back into \eq{Leffapiz} we find that in order 
to renormalize $\Sigma_{\pi\pi}$ and $\Sigma_{a \pi}$ we need to set 
\begin{equation}
   \gamma_3 = -\frac{1}{2}\, , \qquad   \gamma_4 = 2\, .
\end{equation}
Thus the renormalized effective Lagrangian becomes
\begin{equation}\label{LeffReno}
    \mathcal{L}_{a-\piz}^{\rm eff} = \frac{1}{2} a\left( p^2-m_a^2\right)a+\frac{1}{2} \piz\left( p^2 - \widetilde{m}_{\pi}^2 \right) \piz - a \left(p^2 3 C_{a\pi} \frac{\tilde{f_{\pi}}}{2 f_a} + \frac{4 \ell_7 (m_d-m_u) m_u m_d \mpi^4}{f_a f_\pi (m_u+m_d)^3}\right)\piz \, ,
\end{equation}
with
\begin{equation}\label{mpiNLO}
       \widetilde{m}_{\pi}^2  = \mpi^2-\frac{\mpi^4 \overline{\ell_3}}{32\pi^2 f_\pi^2}-\frac{2 \ell_7 (m_d-m_u)^2 \mpi^4}{f_\pi^2(m_u+m_d)^2}\, ,
\end{equation}
and
\begin{equation}\label{fpiNlO}
       \tilde{f}_\pi = f_\pi + \frac{\overline{\ell_4} \mpi^2}{16\pi^2 f_\pi}\, .
\end{equation}
We observe that $\ell_7$ is not renormalized, since in the LO Lagrangian the $m_d-m_u$ terms are all momentum dependent.
So we are left with a non-zero off-diagonal two-point function.
In order to eliminate the mixing, we can rotate the axion and the pion fields as
\begin{equation}
\begin{split}
a &\rightarrow a -\beta_1 \piz\, , \\
\piz &\rightarrow \piz -\beta_2 a \, ,
\end{split}
\end{equation}
yielding
\begin{equation}
\begin{split}
    \mathcal{L}_{a-\piz}^{\rm eff} \rightarrow &\ \frac{1}{2} a\left( p^2-m_a^2\right)a+\frac{1}{2} \piz \left( p^2 - \tilde{m}_{\pi}^2 \right) \piz \\ 
    &- a \left(\beta_1 (p^2-m_a^2) + \beta_2 (p^2-\mpi^2)+p^2 3 C_{a\pi}\frac{\tilde{f_{\pi}}}{2 f_a}+ \frac{4 \ell_7 (m_d-m_u) m_u m_d \mpi^4}{f_a f_\pi (m_u+m_d)^3}\right)\piz \, .
\end{split}
\end{equation}
Hence, to cancel the mixing term it is sufficient to set
\begin{align}
        \beta_1&=-3 C_{a\pi}\frac{\tilde{f_{\pi}}}{2 f_a} -\left(m_\pi^2-m_a^2 \right)^{-1}\left[3 C_{a\pi}\frac{\tilde{f_{\pi}}}{2 f_a} m_a^2+\frac{4 \ell_7 (m_d-m_u) m_u m_d \mpi^4}{f_a f_\pi (m_u+m_d)^3}\right]\, , \\
        \beta_2&= \left(m_\pi^2-m_a^2 \right)^{-1}\left[3 C_{a\pi}\frac{\tilde{f_{\pi}}}{2 f_a} m_a^2 +\frac{4 \ell_7 (m_d-m_u) m_u m_d \mpi^4}{f_a f_\pi (m_u+m_d)^3}\right]\, .
\end{align}

\section{ALP decay amplitudes}
\label{sec:ALPdecayampl}

Following \eq{eq:DecayAmp}, the full ALP decay amplitudes up to NLO 
are given by (employing the definition $\sigma(x)=(1-4m_\pi^2/x)^{1/2}$)
\begin{align} \label{eq:amp1}
&\M_{a\to\piz\pip\pim}=\frac{3  C_{a \pi} m_{\pi }^2 \left(m_{\pi }^2-s\right)}{2 f_a f_{\pi } \left(m_a^2-m_{\pi }^2\right)} \nonumber \\
&+\frac{ C_{a \pi} }{32 \pi^2 f_a f_{\pi }^3  \(m_a^2-\mpi^2\)}\Bigg\{\bar{\ell_1} \mpi^2  \left(2 \mpi^2-s\right) \left(m_a^2+\mpi^2-s\right)\nonumber \\
&+\bar{\ell_2} \mpi^2 \left(m_a^2 \left(2 \mpi^2-s\right)+m_a^4-3 \mpi^2 s+5 \mpi^4-u^2-t^2\right)\nonumber\\
&+\frac{3}{2}\bar{\ell_3} m_{\pi}^4 \frac{m_a^4 - \mpi^2(m_a^2+\mpi^2-s)}{m_a^2-\mpi^2} 
%&+\frac{3 \bar{\ell_3} \mpi^4  \left(\mpi^2 \left(-m_a^2-\mpi^2+s\right)+m_a^4\right)}{2  \left(m_a^2-\mpi^2\right)}\nonumber\\
+3 \bar{\ell_4} \mpi^2  \left(m_a^2+\mpi^2\right) \left(\mpi^2-s\right)\nonumber \\
&+\frac{1}{6}\mpi^2 (m_a^2 \left(45 m_{\pi}^2-29 s\right)+11 m_a^4-15 \mpi^2 s+45 \mpi^4-11 t^2-8 t u-11 u^2)\nonumber\\
%&+\frac{1}{6(m_a^2-\mpi^2)}\Big(m_{\pi}^4 \left(14 m_a^2 s+34 m_a^4+11 t^2+8t u+11 u^2\right)\nonumber\\
%&+m_{\pi}^2 \left(-29 m_a^4 s+m_a^2 \left(-11 t^2-8 t u-11 u^2\right)+11 m_a^6\right)+15 m_{\pi}^6 s-45 m_{\pi}^8\Big)\nonumber\\
&-\frac{1}{2}\log \left(\frac{ \sigma (u)-1}{\sigma (u)+1}\right)\sigma (u)\Big(3 m_{\pi }^4+(u-4 t) m_{\pi }^2\nonumber\\
&+(t-u) u+m_a^2 \left(u-m_{\pi }^2\right)\Big) m_{\pi }^2\nonumber\\
&-\frac{1}{2}\log \left(\frac{ \sigma (t)-1}{\sigma (t)+1}\right)\sigma (t)\Big(3 m_{\pi }^4+(t-4 u) m_{\pi }^2\nonumber\\
&+(u-t) t+m_a^2 \left(t-m_{\pi }^2\right)\Big) m_{\pi }^2\nonumber\\
&+\frac{3 }{2} \log \left(\frac{ \sigma (s)-1}{\sigma (s)+1}\right)\sigma (s)\left(m_{\pi }^2-s\right)  \left(m_a^2+s\right)m_{\pi }^2 \Bigg\}\nonumber\\
&+\ell_7 \frac{\left(m_d-m_u\right) \mpi^4}{3 f_a f_\pi^3 ( m_a^2-m_{\pi }^2) ^2\left(m_d+m_u \right)^3} \Big[\Big(m_a^2 \left(3 s (m_d^2+m_u^2  -6  m_d m_u)-4 \mpi^2 \left(m_d^2 -11  m_d m_u+m_u^2\right)\right)\nonumber\\
&+m_a^4 \left(m_d^2 -14 m_d m_u+m_u^2\right) - 12 m_d \mpi^2 m_u \left(2 \mpi^2-s\right)\Big) \Big] \, ,
\end{align}
and
\begin{align} \label{eq:amp2}
&\M_{a\to\piz\piz\piz}=-\frac{3 C_{a\pi} m_{\pi }^2 m_{{a}}^2}{2 f_{\pi } f_a \left(m_{{a}}^2-m_{\pi }^2\right)} \nonumber \\
&+\frac{ C_{a \pi} }{32 \pi^2 f_a f_{\pi }^3  \(m_a^2-\mpi^2\)}\Bigg\{ 2\bar{\ell_1} \mpi^2  \left(m_a^2 \left(3 \mpi^2-s\right)+m_a^4-3 \mpi^2 s+6 \mpi^4-t^2-t u-u^2\right) \nonumber \\
&+4 \bar{\ell_2}\mpi^2  \left(m_a^2 \left(3 \mpi^2-s\right)+m_a^4-3 \mpi^2 s+6 \mpi^4-t^2-t u-u^2\right)\nonumber\\
%&+\frac{3 \bar{\ell_3}m_a^2 \mpi^4  \left(3 m_a^2-2 \mpi^2\right)}{2(m_a^2-m_{\pi}^2)}
&+\frac{3}{2} \bar{\ell_3}m_a^2 \mpi^4 \frac{3 m_a^2 - 2 \mpi^2}{m_a^2 -  \mpi^2}
-3 \bar{\ell_4}m_a^2 \mpi^2  \left(m_a^2+\mpi^2\right)\nonumber\\
&+\frac{3}{2} \log  \left(\frac{ \sigma (s)-1}{\sigma (s)+1}\right)m_{\pi }^2 \sigma (s)\left(m_{\pi }^2 m_{{a}}^2+2 \left(m_{\pi }^2-s\right)^2\right)\nonumber\\
&+\frac{3}{2} \log  \left(\frac{ \sigma (t)-1}{\sigma (t)+1}\right)m_{\pi }^2 \sigma (t)\left(m_{\pi }^2 m_{{a}}^2+2 \left(m_{\pi }^2-t\right)^2\right)\nonumber\\
&+\frac{3}{2} \log  \left(\frac{ \sigma (u)-1}{\sigma (u)+1}\right)m_{\pi }^2 \sigma (u)\left(m_{\pi }^2 m_{{a}}^2+2 \left(m_{\pi }^2-u\right)^2\right)\nonumber\\
&+\frac{3}{2} \mpi^2  \left(-4 s \left(m_a^2+3 \mpi^2\right)+13 m_a^2 \mpi^2+2 m_a^4+24 \mpi^4-4 \left(t^2+t u+u^2\right)\right)\Bigg\}\nonumber\\
& +\frac{\ell_7\mpi^4 \left(4 m_a^2-3 \mpi^2\right) (m_{d}-m_{u}) \left(m_a^2 (m_{d}^2+m_{u}^2-6 m_d m_u)+4 m_{d} \mpi^2 m_{u}\right)}{f_af_\pi^3 \left(m_a^2-\mpi^2\right)^2 (m_{d}+m_{u})^3} 	 \, .
\end{align}

\bibliographystyle{utphys.bst}

\bibliography{bibliography}

\end{small}

\clearpage

\end{document}